\documentclass[11pt]{article}

\usepackage{multirow}
\usepackage{xcolor}

\usepackage[english]{babel}

\usepackage{blindtext}
\usepackage{tikz}
\usetikzlibrary{backgrounds}

\makeatletter

\newcommand{\Rmnum}[1]{\expandafter\@slowromancap\romannumeral #1@}
\makeatother

\usepackage[linesnumbered,boxed]{algorithm2e}
\usepackage{mathrsfs}
\usepackage{latexsym,lineno}
\usepackage{epsfig}
\usepackage{amsmath}\usepackage{fleqn}\usepackage{verbatim}\usepackage{epsf}
\usepackage{amsthm}\usepackage{graphicx, float}\usepackage{graphicx}
\usepackage{amsfonts}\usepackage{amssymb}\usepackage{graphpap}
\usepackage{epic}\usepackage{curves}

\numberwithin{equation}{section}

\title{\bf A two-component generalization of Burgers equation with Quasi-periodic solutions
}
\author{  Hongfei Pan  ,\, Tiecheng Xia \footnote{{\it Corresponding
author.} E-mail address: xiatc@shu.edu.cn(T. Xia).} ,\, Dengyuan Chen \\
{\small \it Department of Mathematics, Shanghai University, Shanghai
200444, China}
\\
}
\date{}
\begin{document}
\maketitle
\begin{abstract}
In this paper, we aim for the theta function representation of
quasi-periodic solution and related crucial quantities for a
two-component generalization of Burgers equation. Our tools include
the theory of algebraic curve, the meromorphic function,
Baker-Akhiezer functions, the Dubrovin-type equations for auxiliary
divisor, with these tools, the explicit representations for above
quantities are
obtained.\\
\vspace{0.5mm}
\textbf{Keywords}: quasi-periodic solutions; theta
function; divisor

\bigskip

\end{abstract}
\section{Introduction}
The most important physical property of solitons is that they are
localized wave packets which survive collisions with other solitons
without change of shape. Also, solitons found numerous applications
in classical and quantum field theory and in connection with optical
communication devices. For a guide to the vast literature on
solitons, see for instance \cite{Novikov-1984,Degasperis-1998}. The
explicit theta function representations of quasi-periodic solutions
(including soliton solutions as special limiting cases) of
integrable equations are new approach to construing solutions of
integrable nonlinear evolution equations, based on inverse spectral
theory and algebro-geometric
methods\cite{Belokolos-1994,Dubrovin-1981,Its-1975,Novikov-1999}.
The construction of all algebro-geometric solutions and their theta
function representation of some key hierarchy in $1+1$-dimensions
associated with continuous and discrete models had been
done\cite{Gesztesy-2003,Gesztesy-2008}. Based on the
nonlinearization technique of Lax pairs and direct method have been
proposed by Cao\cite{Cao-1990}, through which algebro-geometric
solutions of soliton equations can be
obtained\cite{Cao-1999,Cao-2002,Geng-2000}.

In this paper, we will construct the quasi-periodic solution of the
following two-component generalization of Burgers equation on the
basis of approaches in \cite{Hou-2013,Xue-2013}:
$$
\begin{array}{ll}
u_{t}  =  2u_{xx} + 4u_{x}v,\\
v_{t}  =  -2v_{xx} - 4uu_{x} + 4vv_{x} - 4u_{x}v,
\end{array}\eqno(1.1)
$$
as a reduction case, taking $u=0$, (1.1) reduces to the burgers
equation $v_{t}=-2v_{xx}+4vv_{x}$.

The paper is organized as follows. In Section 2, we introduce the
Lenard gradient to derive a hierarchy, in which the second system is
our equation (1.1). In section 3, we establish a direct relation
between the elliptic variables and the potentials. In section 4, the
hyperelliptic Riemann surface of arithmetic genus $N$ and the
Abel-Jacobi coordinates are introduced from which the corresponding
flows are straighted. Finally, quasi-periodic solution of (1.1) is
given in term of the Riemann theta function, according to the
asymptotic properties and the algebro-geometric characters of the
meromorphic function $\phi$, the Baker-Akhiezer function $\psi_{1}$
and the hyperelliptic curve $\mathcal{K}_{N}$.

\section{The hierarchy and Lax pairs of the two-component generalization of Burgers equation }

In this section, we shall derive a hierarchy of (1.1). Let's
introduce the Lenard gradient sequence $ \{S_{j}\}_{ j = 0,1,2 \dots
}$ by the recursion relation
$$KS_{j-1}= JS_{j},\,\,\ j= 1,2,3, \ldots .\,\,\ S_{j}\arrowvert_{(u,v)=0}=0, \,\,\ S_{0}= (2,-2u,1)^{T}, \eqno(2.1)$$
where $S_{j}=(S_{j}^{(1)}, S_{j}^{(2)}, S_{j}^{(3)})$ and $K, J$ are
two operators defined by($\partial = \partial/\partial x$):
$$
\begin{array}{ccc}
{K}= \left(
\begin{array}{ccc} \partial -(u+v) & 0  &  0 \\
 0 & \partial + (u+v) & 0  \\
 -u & -1& \partial \
\end{array}\right)
\end{array},
\begin{array}{ccc}
{J}= \left(
\begin{array}{ccc} 1  &  0 & -2 \\
 0 & -1 & -2u \\
 -u & -1 & \partial\
\end{array}\right)
\end{array}. \eqno(2.2)
$$
A direct calculation gives from the recursion
relation(2.1)that$$S_{1}= (2(u-v), 2u_{x}+2u(v-u),
2u)^{T},\eqno(2.3)$$
$$S_{2}=\begin{array}{ccc}
\left(
\begin{array}{ccc}
2u^2 - 2u_{x}-2v_{x}-8uv+2v^2 \\
-2u^3 + 6uu_{x} + 8u^{2}v - 2uv^2 -2u_{xx} -4u_{x}v - 2uv_{x} \\
2u^2 - 2u_{x} -4uv \
\end{array}\right)
\end{array}.\eqno(2.4)$$
Consider the spectral problem :
$$\psi_{x}= U \psi , \,\,\ \begin{array}{ccc}
{U}= \left(
\begin{array}{cc} -\frac{1}{2}(\lambda^2 + u + v) & \lambda u \\
  - \lambda  & \frac{1}{2}(\lambda^2 + u + v) \
\end{array}\right)
\end{array}, \eqno(2.5)$$
and the auxiliary problem:
$$\psi_{t_{m}}= V^{(m)} \psi , \,\,\ \begin{array}{ccc}
{V^{(m)}}= \left(
\begin{array}{cc} V_{11}^{(m)} & V_{12}^{(m)} \\
 V_{21}^{(m)} & - V_{11}^{(m)} \
\end{array}\right)
\end{array}, \eqno(2.6)$$
where $V_{11}^{(m)}, V_{12}^{(m)}, V_{21}^{(m)}$ are polynomials of
the spectral parameter $\lambda$ with:
$$
\begin{array}{ll}
&V_{11}^{(m)}= \frac{1}{2}S_{mx}^{(1)}+\frac{1}{2}(u+v)S_{m}^{(1)}+
\sum\limits_{j=0}^m S_{j}^{(3)}\lambda^{2(m-j)+2},\\& V_{12}^{(m)}=
\sum\limits_{j=0}^m S_{j}^{(2)}\lambda^{2(m-j)+1}, \\&V_{21}^{(m)}=
\sum\limits_{j=0}^m S_{j}^{(1)}\lambda^{2(m-j)+1}.\end{array}
\eqno(2.7)$$ Then the compatibility condition of (2.5) and (2.6)
yields the zero curvature equation, $U_{t_{m}}- V_{x}^{(m)}+
[U,V^{(m)}]=0,$ which is equivalent to a hierarchy of nonlinear
evolution equations
$$
\begin{array}{ll}
&u_{t_{m}} =  S_{mx}^{(2)} + (u+v)S_{m}^{(2)} + uS_{mx}^{(1)} +
u(u+v)S_{m}^{(1)},\\& v_{t_{m}} =  S_{mxx}^{(1)} - \partial
(u+v)S_{m}^{(1)}- S_{mx}^{(2)}-(u+v)S_{m}^{(2)}-uS_{mx}^{(1)} -
u(u+v)S_{m}^{(1)}. \end{array} \eqno(2.8)$$ The second equations in
(2.8) is $(m=1)$
$$
\begin{array}{ll}
u_{t_{1}}  =  2u_{xx} + 4u_{x}v,\\
v_{t_{1}}  =  -2v_{xx} - 4uu_{x} + 4vv_{x} - 4u_{x}v,
\end{array}
$$
and it is our equation (1.1).

\section{Evolution of elliptic variables}
In this section, we shall establish a relation between the elliptic
variables and the potentials. Let $\psi = (\psi_{1}, \psi_{2})^{T}$
and $\phi = (\phi_{1}, \phi_{2})^{T} $ be two basic solutions of
(2.5) and (2.6), We define a matrix W by
$$W= \frac{1}{2}(\phi \psi^{T} + \psi \phi^{T})\begin{array}{ccc}
\left(
\begin{array}{cc} 0 & -1 \\
 1 & 0 \
\end{array}\right) = \begin{array}{ccc}
\left(
\begin{array}{cc}G & F \\
 H & -G \
\end{array}\right)
\end{array}
\end{array}. \eqno(3.1)$$
It is easy to calculate by (2.5) and (2.6) that
$$W_{x} = [U,W],\,\,\ W_{t_{m}}=[V^{m},W], \eqno(3.2)$$
which implies that $\partial_{x} \ detW = 0, \,\,\
\partial_{t_{m}} \ detW = 0.$ Equation (3.2) can be written as
$$
\begin{array}{ll}
G_{x} = \lambda(uH + F),\\
F_{x} = -F \lambda^2 -(u+v)F - 2uG \lambda ,\\
H_{x}= H\lambda^2 + (u+v)H - 2G \lambda,
\end{array}\eqno(3.3)
$$
and
$$\begin{array}{ll}
G_{t_{m}}& = HV_{12}^{(m)} - FV_{21}^{(m)},\\
F_{t_{m}}& = 2FV_{11}^{(m)} - 2GV_{12}^{(m)},\\
H_{t_{m}}&= 2GV_{21}^{(m)} - 2HV_{11}^{(m)}.
\end{array}\eqno(3.4)$$
We suppose that the functions $G, F, H$ are finite-order polynomials
in $\lambda$:
$$G = \sum\limits_{j=0}^{N}g_{2j+1}\lambda^{2(N-j)+1}, \,\ F = \sum\limits_{j=0}^{N}f_{2j}\lambda^{2(N-j)}, \,\
H = \sum\limits_{j=0}^{N}h_{2j}\lambda^{2(N-j)}.\eqno(3.5)$$
Substituting (3.5) into (3.3) yields:
$$KG_{j-1}= JG_{j} (j=1,2,\ldots,N), \,\,\ JG_{0}=0, \,\,\ KG_{N}=0, \,\,\
G_{j}=(h_{2j},f_{2j},g_{2j+1})^{T}.\eqno(3.6)$$  The equation
$JG_{0}=0$ has the general solution
$$G_{0}=\alpha_{0}S_{0}, \eqno(3.7)$$ where $\alpha_{0}$ is a constant
of integration, and let $\alpha_{0}=1$ without loss of generality.
If we take (3.7) as a starting point, then $G_{j}$ can be
recursively determined by the relation(3.6). In fact, noticing $\ker
J = \{cS_{0}|\forall c \in \mathbb{C} \}$ and acting with the
operator $(J^{-1}K)^k$ upon (3.7), we obtain from (3.6) and (2.1)
that:
$$G_{k}=\sum\limits_{j=0}^{k}\alpha_{j}S_{k-j}, \,\ k=0,1,\ldots , N , \eqno(3.8)$$
where $\alpha_{1}, \ldots ,\alpha_{k}$ are integral constants. The
first few members in (3.8) are:
$$
\begin{array}{ll}
&G_{0}= \begin{array}{ccc} \left(
\begin{array}{ccc}
2 \\
-2u \\
1 \
\end{array}\right)
\end{array},
G_{1}= \begin{array}{ccc} \left(
\begin{array}{ccc}
2(u-v) \\
2u_{x}+2u(v-u) \\
 2u \
\end{array}\right) + \alpha_{1} \left(
\begin{array}{ccc}
2 \\
-2u \\
1 \
\end{array}\right)
\end{array},\\&
\begin{array}{ll} G_{2}= \left(
\begin{array}{ll}
2u^2-2u_{x}-2v_{x}-8uv+2v^2 \\
-2u^3+6uu_{x}+8u^2v-2uv^2-2u_{xx}-4u_{x}v-2uv_{x} \\
2u^2-2u_{x}-4uv \
\end{array}\right)
\end{array} \\&
 \,\,\,\,\,\,\,\,\,\,\,\,\,\,\,\,\   + \alpha_{1}\begin{array}{ll} \left(
\begin{array}{ll}
2(u-v) \\
2u_{x}+2u(v-u) \\
2u \
\end{array}\right)
\end{array}
+ \alpha_{2} \begin{array}{ll} \left(
\begin{array}{ll}
2 \\
-2u \\
1 \
\end{array}\right)
\end{array}.
\end{array}
\eqno(3.9)
$$
We write $F$ and $H$ as the following finite products:
$$\begin{array}{ll}
F = -2u\prod\limits_{j=1}^N (\lambda^2 -
u_{j}^2):=-2u\prod_{j=1}^N(\widetilde{\lambda}-\widetilde{u}_{j}),
\,\,\ H = 2 \prod\limits_{j=1}^N (\lambda^2 -
v_{j}^2):=2\prod\limits_{j=1}^N(\widetilde{\lambda}-\widetilde{v}_{j}),
\end{array}
\eqno(3.10)$$ where $\widetilde{\lambda}=\lambda^2,
\widetilde{u}_{j}=u_{j}^2, \widetilde{v}_{j}=v_{j}^2$, and
$\{\widetilde{u}_{j}\}_{j=1, \ldots, N }$ and
$\{\widetilde{v}_{j}\}_{j=1, \ldots, N}$ are called elliptic
variables, comparing the coefficients of $\widetilde{\lambda}^{N-1}$
in the expressions for $F$ and $H$ in (3.5) and (3.10),
respectively, we obtain:
$$\partial\ln u=\sum\limits_{j=1}^N (\widetilde{u}_{j}-\widetilde{v}_{j}), \eqno(3.11)$$
$$v-u=\sum\limits_{j=1}^N \widetilde{v}_{j} + \alpha_{1}. \eqno(3.12)$$
Similarly, comparing the coefficients of $\widetilde{\lambda}^{N-2}$
and $\widetilde{\lambda}^{0}$ in (3.5) and (3.10), respectively, we
have:
$$u^2-3u_{x}-4uv+v^2 + \frac{u_{xx}}{u} + \frac{2u_{x}v}{u} + v_{x} -
\alpha_{1}[\partial\ln u +(v-u)] + \alpha_{2} = \sum\limits_{j < k}
\widetilde{u}_{j}\widetilde{u}_{k}, \eqno(3.13)$$
$$u^2 - u_{x} -v_{x} - 4uv + v^2 + \alpha_{1}(u-v) + \alpha_{2} = \sum\limits_{j < k}
\widetilde{v}_{j}\widetilde{v}_{k}, \eqno(3.14) $$
$$f_{2N} = (-1)^{N+1}2u\prod\limits_{j=1}^{N}\widetilde{u}_{j}, \,\,\
h_{2N} = (-1)^{N}2\prod\limits_{j=1}^{N}\widetilde{v}_{j}.
\eqno(3.15)$$ Since $\det W$ is a $(2N+1)$th-order polynomial in
$\widetilde{\lambda}$ with constant coefficients of the $x$- flow
and $t_{m}$-flow, we have:
$$-\det W=G^{2} + FH = \prod_{j=1}^{2N+1}(\lambda^2 - \lambda_{j}^2)
=\prod_{j=1}^{2N+1}(\widetilde{\lambda}-\widetilde{\lambda}_{j}) =
\frac{1}{\widetilde{\lambda}} R(\widetilde{\lambda}),\eqno(3.16)$$
from which we obtain
$$G|_{\widetilde{\lambda}=\widetilde{u}_{k}}=\sqrt{\frac{R(\widetilde{u}_{k})}{\widetilde{u}_{k}}},
\,\,\ G|_{\widetilde{\lambda} = \widetilde{v}_{k}} =
\sqrt{\frac{R(\widetilde{v}_{k})}{\widetilde{v}_{k}}}. \eqno(3.17)$$
By using (3.3) and (3.10), we get:
$$F_{x}|_{\widetilde{\lambda} = \widetilde{u}_{k}} = 2u\widetilde{u}_{k, x}\prod_{j=1,j\neq k}^N (\widetilde{u}_{k} - \widetilde{u}_{j})=
-2u\sqrt{\widetilde{u}_{k}}G|_{\widetilde{\lambda}=
\widetilde{u}_{k}}, \eqno(3.18)$$
$$H_{x}|_{\widetilde{\lambda} = \widetilde{v}_{k}} = -2\widetilde{v}_{k, x}\prod_{j=1,j\neq k}^N (\widetilde{v}_{k}-\widetilde{v}_{j})=
-2\sqrt{\widetilde{v}_{k}}G|_{\widetilde{\lambda} =
\widetilde{v}_{k}}, \eqno(3.19)$$ which means:
$$\widetilde{u}_{k, x}= \frac{-\sqrt{R(\widetilde{u}_{k})}}{\prod\limits_{j=1,j\neq
k}^N (\widetilde{u}_{k} - \widetilde{u}_{j})}, v_{k,x}=
\frac{\sqrt{R(\widetilde{v}_{k})}}{\prod\limits_{j=1,j\neq k}^N
(\widetilde{v}_{k} - \widetilde{v}_{j})},\,\,\,\ 1 \leq k \leq N.
\eqno(3.20)$$ In a way similar to the above expression , by using
(3.4), (3.10), (3.17), we get the evolution of
$\{\widetilde{u}_{k}\}$ and $\{\widetilde{v}_{k}\}$ along the
$t_{m}-$ flow:
$$\widetilde{u}_{k,t_{m}}= -\frac{G|_{\widetilde{\lambda} = \widetilde{u}_{k}}V_{12}^{(m)}|_{\widetilde{\lambda} = \widetilde{u}_{k}}}
{u \prod\limits_{j=1,j\neq k}^N (\widetilde{u}_{k} -
\widetilde{u}_{j})}, \widetilde{v}_{k,t_{m}} = -
\frac{G|_{\widetilde{\lambda} =
\widetilde{v}_{k}}V_{21}^{(m)}|_{\widetilde{\lambda}=\widetilde{v}_{k}}}
{\prod\limits_{j=1,j\neq k}^N
(\widetilde{v}_{k}-\widetilde{v}_{j})}. \eqno(3.21)
$$ When $m=1$, associate with (3.11) and (3.12) and set
$t_{1}=t$, we have:
$$\widetilde{u}_{k, t}= \frac{2\sqrt{R(\widetilde{u}_{k})}(\widetilde{u}_{k} - \sum\limits_{j=1}^N \widetilde{u}_{j} - \alpha_{1})}
{\prod\limits_{j=1,j\neq k}^N
(\widetilde{u}_{k}-\widetilde{u}_{j})}, \,\,\ \widetilde{v}_{k,t} =
\frac{2\sqrt{R(\widetilde{v}_{k})}(-\widetilde{v}_{k}+
\sum\limits_{j=1}^N \widetilde{v}_{j} + \alpha_{1})}
{\prod\limits_{j=1,j\neq k}^N (\widetilde{v}_{k} -
\widetilde{v}_{j})}. \eqno(3.22)$$

\section{Quasi-periodic solution}
In this section, we shall construct quasi-periodic solution of
(1.1). Noticing (3.16), we introduce the hyperelliptic curve
$\mathcal{K}_{N}$ of arithmetic genus $N$ defined by:
$$\mathcal{K}_{N}: \,\,\,\ y^2 - R(\widetilde{\lambda}) = 0, \,\ R(\widetilde{\lambda})=
\prod\limits_{j=1}^{2N+2}(\widetilde{\lambda}-\widetilde{\lambda}_{j}),
\,\ \widetilde{\lambda}_{2N+2}=0. \eqno(4.1)$$ The curve
$\mathcal{K}_{N}$ can be compactified by joining two points at
infinity $P_{\infty_{\pm}}, \,\ P_{\infty_{+}} =
(P_{\infty_{-}})^{\ast}$. For notational simplicity the
compactification is also denoted by $\mathcal{K}_{N}$. Here we
assume that the zeros $\widetilde{\lambda}_{j}, \,\ j=1, \ldots,
2N+2$ of $R(\widetilde{\lambda})$ in (4.1) are mutually distinct,
then the hyperelliptic curve $\mathcal{K}_{N}$ becomes nonsingular.
According to the definition of $\mathcal{K}_{N}$, we can lift the
roots $\{\widetilde{u}_{j}\}_{j=1, \ldots, N}, \,\,\
\{\widetilde{v}_{j}\}_{j=1, \ldots, N}$ to $\mathcal{K}_{N}$ by
introducing:
$$\widehat{\widetilde{u}}_{j}(x, t_{m})=(\widetilde{u}_{j}(x, t_{m}),
\widetilde{\lambda}^{\frac{1}{2}}G(\widetilde{u}_{j}(x, t_{m}))),
\,\,\ j=1, \ldots, N, \eqno(4.2)$$
$$\widehat{\widetilde{v}}_{j}(x, t_{m})=(\widetilde{v}_{j}(x, t_{m}),
-\widetilde{\lambda}^{\frac{1}{2}}G(\widetilde{v}_{j}(x, t_{m}))),
\,\,\ j=1, \ldots, N, \eqno(4.3)$$ where $j=1, \ldots, N, \,\
(x,t_{m})\in \mathbb{R}^2$.

We give the definition of the meromorphic function $\phi(\cdot, x,
t_{m})$ on $\mathcal{K}_{N}$ by (4.1) and (3.16)
$$\phi(P, x, t_{m}) = \frac{y + \widetilde{\lambda}^{\frac{1}{2}}G}{\widetilde{\lambda}^{\frac{1}{2}}F}
= \frac{\widetilde{\lambda}^{\frac{1}{2}}H}{y -
\widetilde{\lambda}^{\frac{1}{2}}G}, \eqno(4.4)$$ where $P =
(\widetilde{\lambda}, y) \in \mathcal{K}_{N} \setminus
 \{P_{\infty_{\pm}}\} $, hence the divisor of
 $\phi(\cdot, x, t_{m})$ reads$\cite{Gesztesy-2003,Griffiths-1994}$
$$(\phi(\cdot, x, t_{m}))= \mathcal{D}_{P_{\infty_{+}}\underline{\widehat{\widetilde{v}}}(x, t_{m})} -
 \mathcal{D}_{P_{\infty_{-}} \underline{\widehat{\widetilde{u}}}(x, t_{m})}, \eqno(4.5)$$
where $$\mathcal{D}_{ \underline{\widehat{\widetilde{u}}}}(x,
t_{m})(P) = \sum\limits_{j=1}^N \widehat{\widetilde{u}}_{j}
(x,t_{m}), \,\ \mathcal{D}_{ \underline{\widehat{\widetilde{v}}}}(x,
t_{m})(P) = \sum\limits_{j=1}^N \widehat{\widetilde{v}}_{j}
(x,t_{m}),$$ and $P_{\infty_{+}}, \,\
\widehat{\widetilde{v}}_{1}(x,t_{m}), \ldots,
\widehat{\widetilde{v}}_{N}(x,t_{m})$ are the $N+1$ zeros of
$\phi(P, x, t_{m})$. $P_{\infty_{-}}, \,\
\widehat{\widetilde{u}}_{1}(x,t_{m}), \ldots,
\widehat{\widetilde{u}}_{N}(x,t_{m})$ are the $N+1$ poles of
$\phi(P, x, t_{m})$.

On the basis of the definition of meromorphic function $\phi(\cdot,
x, t_{m})$ in (4.4), the spectral problem (2.5) and the auxiliary
problem (2.6), we can define the Baker-Akhiezer vector $\Psi(\cdot,
x, x_{0}, t_{m}, t_{m,0})$ on $\mathcal{K}_{N} \backslash
\{P_{\infty_{+}}, P_{\infty_{-}} \}$ by
$$\Psi(P,
x, x_{0}, t_{m}, t_{m,0}) = \begin{array}{ccc} \left(
\begin{array}{ccc}
 \psi_{1}(P, x, x_{0}, t_{m}, t_{m,0})\\
\psi_{2}(P, x, x_{0}, t_{m}, t_{m,0}) \
\end{array}\right)
\end{array}, \eqno(4.6)$$
where
$$\begin{array}{ll} \psi_{1}(P, x, x_{0}, t_{m}, t_{m,0}) =
&exp(\int_{x_{0}}^{x}(-\frac{1}{2})(\widetilde{\lambda} +
u(x^{'},t_{m})+ v(x^{'}, t_{m})) -
\sqrt{\widetilde{\lambda}}u(x^{'}, t_{m})\phi(P,x^{'},t_{m}))dx^{'}
\\ & + \int_{t_{m,0}}^{t_{m}}(V_{11}^{(m)}(\widetilde{\lambda}, x_{0}, s) -
V_{12}^{(m)}(\widetilde{\lambda}, x_{0}, s)\phi(P, x_{0}, s))ds)
\end{array}, \eqno(4.7)$$
$$\psi_{2}(P,x,x_{0},t_{m},t_{m,0}) = -\psi_{1}(P,x,x_{0},t_{m},t_{m,0})\phi(P,x,t_{m}), \eqno(4.8)$$
with $P \in \mathcal{K}_{N} \backslash \{P_{\infty_{+}},
P_{\infty_{-}} \} , (x,t_{m}), (x_{0}, t_{m,0}) \in \mathbb{R}^2$.

Next, we introduce the Riemann surface $\Gamma$ of the hyperelliptic
curve $\mathcal{K}_{N}$ and equip $\Gamma$ with a canonical basis of
cycles: $a_{1}, a_{2},\ldots, a_{N}; b_{1}, b_{2}, \ldots, b_{N}$
which are independent and have intersection numbers as follows:
$$a_{i}\circ a_{j}=0, \,\,\
b_{i}\circ b_{j}=0, \,\,\ a_{i}\circ b_{j}=\delta_{ij},
i,j=1,2,\ldots,N.$$

We choose the following set as our basis:
$$\widetilde{\omega}_{l}=\frac{\widetilde{\lambda}^{l-1}d\widetilde{\lambda}}{\sqrt{R(\widetilde{\lambda})}},
\,\,\ l=1,2,\ldots,N,$$ which are linearly independent homomorphic
differentials from each other on $\Gamma$, and let
$$A_{ij}=\int_{a_{j}}\widetilde{\omega}_{i}, \,\,\
B_{ij}=\int_{b_{j}}\widetilde{\omega}_{i}.$$ It is possible to show
that the matrices $A=(A_{ij})$ and $B=(B_{ij})$ are $N \times N$
invertible period matrices \cite{Siegel-1971,Griffiths-1994}. Now we
define the matrices $C$ and $\tau$ by $C=(C_{ij}) = A^{-1}, \,\,\
\tau=(\tau_{ij}) = A^{-1}B$. Then the matrix $\tau$ can be shown to
symmetric $(\tau_{ij}=\tau_{ji})$ and it has a positive-definite
imaginary part (Im $\tau > 0$). If we normalize
$\widetilde{\omega}_{j}$ into the new basis $\omega_{j}$:
$$\omega_{j}= \sum_{l=1}^N C_{jl}\widetilde{\omega}_{l}, \,\,\,\,\ l=1,2, \ldots , N,$$
then we have:
$$\int_{a_{j}}\omega_{j} = \sum_{l=1}^N C_{jl}\int_{a_{j}}\widetilde{\omega}_{l} = \sum_{l=1}^N C_{jl}A_{li} = \delta_{ji},$$
$$\int_{b_{j}}\omega_{i} = \sum_{l=1}^N C_{jl}\int_{b_{j}}\widetilde{\omega}_{l} = \sum_{l=1}^N C_{jl}B_{li} = \tau_{ji}.$$
Now we define the Abel-Jacobi coordinates:
$$\rho_{j}^{(1)}(x, t_{m}) = \sum_{k=1}^N \int_{P_{0}}^{\widehat{\widetilde{u}}_{k}(x, t_{m})}\omega_{j}=
\sum_{k=1}^N\sum_{l=1}^N\int_{\lambda(P_{0})}^{\widetilde{u}_{k}(x,
t_{m})}C_{jl}
\frac{\widetilde{\lambda}^{l-1}d\widetilde{\lambda}}{\sqrt{R(\widetilde{\lambda})}},
\eqno(4.9)$$
$$\rho_{j}^{(2)}(x,t_{m}) = \sum_{k=1}^N \int_{P_{0}}^{\widehat{\widetilde{v}}_{k}(x, t_{m})}\omega_{j}=
\sum_{k=1}^N\sum_{l=1}^N\int_{\lambda(P_{0})}^{\widetilde{v}_{k}(x,
t_{m})}C_{jl} \frac{\widetilde{\lambda}^{l-1}d
\widetilde{\lambda}}{\sqrt{R(\widetilde{\lambda})}}, \eqno(4.10)$$
where $\lambda(P_{0})$ is the local coordinate of $P_{0}$. From
(3.20) and (4.9), we get
$$\partial_{x}\rho_{j}^{(1)}=
\sum_{k=1}^N \sum_{l=1}^N
C_{jl}\frac{\widetilde{u}_{k}^{l-1}\widetilde{u}_{k,
x}}{\sqrt{R(\widetilde{u}_{k})}}= -\sum_{k=1}^N\sum_{l=1}^N
\frac{C_{jl}\widetilde{u}_{k}^{l-1}}{\prod\limits_{j=1,j\neq
k}^N(\widetilde{u}_{k}-\widetilde{u}_{j})},$$ which implies
$$\partial_{x}\rho_{j}^{(1)}= -C_{jN}= \Omega_{j}^{(1)}, \,\,\,\
j=1,2,\ldots, N .\eqno(4.11)$$ With the help of the following
equality:
$$
\sum_{k=1}^N\frac{u_{k}^{l-1}}{\prod\limits_{i=1,i \neq
k}^N(u_{k}-u_{i})}=\left \{
\begin{array}{ll}
\delta_{lN} ,& \textrm{$l= 1,2,\ldots,N,$}
\\ \sum \limits_{i_{1}+i_{2}+\ldots
+i_{N}=l-N}u_{1}^{i_{1}}u_{2}^{i_{2}} \cdots u_{N}^{i_{N}},&
\textrm{$ l > N $} .
\end{array}
\right.$$

In a similar way, we obtain from (4.9), (4.10), (3.20), (3.21):
$$\partial_{t}\rho_{j}^{(1)} = 2C_{j,N-1} - 2\alpha_{1}C_{j,N} =
 \Omega_{j}^{(2)},  \,\,\,\ j=1,2,\ldots ,N, \eqno(4.12)$$
$$\partial_{x}\rho_{j}^{(2)}=-\Omega_{j}^{(1)}, \,\,\,\ j=1,2,\ldots ,N, \eqno(4.13)$$
$$\partial_{t}\rho_{j}^{(2)}=-\Omega_{j}^{(2)}, \,\,\,\ j=1,2,\ldots ,N.\eqno(4.14)$$

Let $\mathcal{T}$ be the lattice generated by $2N$ vectors
$\delta_{j}, \tau_{j}$, where $\delta_{j} = (\underbrace{0, \ldots ,
0}\limits_{j-1}, 1, \underbrace{0, \ldots, 0}\limits_{N-j})$ and
$\tau_{j}= \tau \delta_{j}$, the Jacobian variety of $\Gamma$ is
$\mathcal{J}= \mathbb{C}^N / \mathcal{T}$. On the basis of these
results, we obtain the following:
$$\rho_{j}^{(1)}(x,t)=\Omega_{j}^{(1)}x + \Omega_{j}^{(2)}t + \gamma_{j}^{(1)}, \eqno(4.15)$$
$$\rho_{j}^{(2)}(x,t)=-\Omega_{j}^{(1)}x - \Omega_{j}^{(2)}t + \gamma_{j}^{(2)}, \eqno(4.16)$$
where $\gamma_{j}^{(i)}(i=1,2)$ are constants, and
$$\underline{\rho}^{(1)} = (\rho_{1}^{((1)}, \rho_{2}^{(1)} \ldots , \rho_{N}^{(1)})^{T}, \,\,\,\
\underline{\rho}^{(2)} = (\rho_{1}^{((2)}, \rho_{2}^{(2)}, \ldots ,
\rho_{N}^{(2)})^{T},$$
$$\underline{\Omega}^{(m)}=(\Omega_{1}^{(m)}, \Omega_{2}^{(m)}, \ldots ,
\Omega_{N}^{(m)})^{T}, \,\,\
\underline{\gamma}^{(m)}=(\gamma_{1}^{(m)}, \gamma_{2}^{(m)}, \ldots
, \gamma_{N}^{(m)})^{T}, \,\,\,\ m=1,2.$$ Now we introduce the Abel
map $\mathcal{A}(P): Div(\Gamma)\rightarrow \mathcal{J}$:
$$\mathcal {A}(P)=\int_{P_{0}}^P \underline{\omega} , \,\,\,\ \underline{\omega}=(\omega_{1}, \omega_{2}, \ldots, \omega_{N})^T,$$
$$\mathcal {A}(\sum_{k}n_{k}P_{k})=\sum n_{k}\mathcal {A}(P_{k}), \,\,\ P, \,\ P_{k} \in \mathcal{K}_{N},$$
the Riemann theta function is defined as
\cite{Gesztesy-2003,Siegel-1971,Griffiths-1994}
$$
\begin{array}{ll} &\theta(P, D_{\widehat{\underline{\widetilde{u}}}(x, t_{m})}) =
\theta(\underline{\Lambda} - \mathcal{A}(P) +
\underline{\rho}^{(1)}) ,\\& \theta(P,
D_{\widehat{\underline{\widetilde{v}}}(x, t_{m})}) =
\theta(\underline{\Lambda} - \mathcal{A}(P) +
\underline{\rho}^{(2)}), \end{array} \eqno(4.17)$$ where
$\underline{\Lambda} = (\Lambda_{1},\ldots , \Lambda_{N})$ is
defined by:
$$\Lambda_{j}=\frac{1}{2}(1 + \tau_{jj}) - \sum\limits_{i=1,i\neq
j}^N \int_{a_{i}} \omega_{i} \int_{Q_{0}}^P \omega_{j}, \,\,\,\ j=1,
\ldots , N.$$

In order to derive the quasi-periodic solution of (1.1), now we turn
to the asymptotic properties of the meromorphic function $\phi$ and
Baker-Akhiezer function $\psi_{1}$.

{\bf Lemma 4.1.} Suppose that $u(x, t_{m}), v(x, t_{m}) \in
C^{\infty}(\mathbb{R}^2)$ satisfy the equation (1.1). Moreover, let
$P=(\widetilde{\lambda}, y) \in \mathcal{K}_{N}\backslash
\{P_{\infty_{\pm}}, P_{0}\}, \,\ (x, x_{0}) \in \mathbb{R}^2$. Then:
$$\begin{array}{ll}
\phi(P) \underset{\zeta \rightarrow 0}{=} \left \{
\begin{array}{ll}
-\zeta^{\frac{1}{2}} -v\zeta^{\frac{3}{2}} + O(\zeta^{\frac{5}{2}})
\,\,\ as \,\,\ P \rightarrow P_{\infty_{+}},
\\  -\frac{1}{u}\zeta^{-\frac{1}{2}} - (\frac{u_{x}}{u^2} + \frac{v}{u})\zeta^{\frac{1}{2}} + O(\zeta^{\frac{3}{2}}) \,\,\ as \,\,\ P
\rightarrow P_{\infty_{-}},
\end{array}
\right.
\end{array}
\eqno(4.18)$$

$$\phi(P)  \underset{\zeta \rightarrow 0}{=} 1 - 2\partial_{t}^{-1}((u+v)(u-v)+(u_{x}-v_{x})) + O(\zeta)
\,\,\,\,\ as \,\,\,\ P \rightarrow P_{0}, \eqno(4.19)$$

$$\begin{array}{ll}
\psi_{1}(P, x, x_{0}, t_{m}, t_{m,0}) \underset{\zeta \rightarrow
0}{=} \left \{
\begin{array}{ll}
exp(-\frac{1}{2}\zeta^{-1}(x-x_{0}) + \zeta^{-m-1}(t_{m} - t_{m,0})
+ O(1)) \,\,\ as \,\,\ P \rightarrow P_{\infty_{+}},
\\ exp(\frac{1}{2}\zeta^{-1}(x-x_{0}) - \zeta^{-m-1}(t_{m} - t_{m,0})
+ O(1)) \,\,\ as \,\,\ P \rightarrow P_{\infty_{-}},
\end{array}
\right.
\end{array}
\eqno(4.20)$$

{\bf Proof.} We first prove $\phi$ satisfies the Riccati-type
equations:
$$\phi_{x}(P) - [\widetilde{\lambda} + (u+v)]\phi(P)- \sqrt{\widetilde{\lambda}}u\phi^2(P)
= \sqrt{\widetilde{\lambda}}, \eqno(4.21)$$
$$\phi_{t}(P) + 2V_{11}^{(1)}\phi(P) - V_{12}^{(1)}\phi^2 (P) = -V_{21}^{(1)}. \eqno(4.22)$$
The local coordinates $\zeta = \widetilde{\lambda}^{-1}$ near
$P_{\infty_{\pm}}$ and $\zeta = \widetilde{\lambda}$ near $P_{0}$,
from (4.4), (3.3),(4.1), we have
$$\phi_{x} = \frac{\widetilde{\lambda}G_{x}F - (y + \widetilde{\lambda}^{\frac{1}{2}}G)\widetilde{\lambda}^{\frac{1}{2}}F_{x}}
{\widetilde{\lambda}F^2} =\widetilde{\lambda}^{\frac{1}{2}} +
[\widetilde{\lambda}+(u+v)]\phi +
\frac{(uH+2uG\phi)\widetilde{\lambda}^{\frac{1}{2}}}{F},
\eqno(4.23)$$
$$\phi^2 = \frac{y^2 + 2\widetilde{\lambda}^{\frac{1}{2}}yG + \widetilde{\lambda}G^2}{\widetilde{\lambda}F^2}
=\frac{2\widetilde{\lambda}G^2 + \widetilde{\lambda}FH +
2\widetilde{\lambda}^{\frac{1}{2}}yG}{\widetilde{\lambda}F^2} =
\frac{2G\phi + H}{F}, \eqno(4.24)$$ according to (4.23) and (4.24),
we have (4.21). Similarly, by using (4.4), (3.4), (4.1), we have
(4.22). And then, inserting the ansatz $\phi
\underset{\widetilde{\lambda}\rightarrow 0}{=}
\phi_{1}\widetilde{\lambda}^{-\frac{1}{2}} +
\phi_{2}\widetilde{\lambda}^{-\frac{3}{2}} +
O(\widetilde{\lambda}^{-\frac{5}{2}})$ into (4.21), we get the first
line of (4.18). Inserting he ansatz $\phi
\underset{\widetilde{\lambda}\rightarrow 0}{=}
\phi_{-1}\widetilde{\lambda}^{\frac{1}{2}} +
\phi_{1}\widetilde{\lambda}^{-\frac{1}{2}} +
O(\widetilde{\lambda}^{-\frac{3}{2}})$ into (4.21), we get the
second line of (4.18). In exactly the same manner, inserting the
ansatz $\phi \underset{\widetilde{\lambda}\rightarrow 0}{=} \phi_{0}
+ \phi_{1}\widetilde{\lambda} + O(\widetilde{\lambda}^{2})$ into
(4.22) immediately yields (4.19).

In the following, we will prove (4.20). From (4.7) and (4.18):
$$\begin{array}{ll}
& \,\,\,\,\,\  exp(\int_{x_{0}}^x
((-\frac{1}{2})(\widetilde{\lambda} + u(x^{'}, t_{m}) + v(x^{'},
t_{m})) - \widetilde{\lambda}^{\frac{1}{2}}u(x^{'}, t_{m})\phi(P,
x^{'}, t_{m}))dx^{'}) \\&= exp(\int_{x_{0}}^x
((-\frac{1}{2})(\zeta^{-1} + u + v) -
\zeta^{-\frac{1}{2}}u\phi)dx^{'}) \\&\underset{\zeta \rightarrow
0}{=} \left \{
\begin{array}{ll}
exp(\int_{x_{0}}^x (-\frac{1}{2})(\zeta^{-1} + u + v) -
\zeta^{-\frac{1}{2}}u(-\zeta^{\frac{1}{2}} -v\zeta^{\frac{3}{2}} +
O(\zeta^{\frac{5}{2}}))) \,\,\ as \,\,\ P \rightarrow
P_{\infty_{+}},
\\ exp(\int_{x_{0}}^x (-\frac{1}{2})(\zeta^{-1} + u + v) -
\zeta^{-\frac{1}{2}}u(-\frac{1}{u}\zeta^{-\frac{1}{2}} -
(\frac{u_{x}}{u^2} + \frac{v}{u})\zeta^{\frac{1}{2}} +
O(\zeta^{\frac{3}{2}})) \,\,\ as \,\,\ P \rightarrow P_{\infty_{-}},
\end{array}
\right.\\& \underset{\zeta \rightarrow 0}{=}\left \{
\begin{array}{ll}
exp(-\frac{1}{2}\zeta^{-1}(x - x_{0}) + O(1)) \,\,\ as \,\,\ P
\rightarrow P_{\infty_{+}},
\\ exp(\frac{1}{2}\zeta^{-1}(x - x_{0}) + O(1) \,\,\ as \,\,\ P \rightarrow
P_{\infty_{-}}.
\end{array}
\right.
\end{array}
\eqno(4.25)$$ From (4.1) and (3.16), we have
$$\begin{array}{ll}
y &= \mp\sqrt{R(\widetilde{\lambda})}\\&
=\mp\widetilde{\lambda}^{\frac{1}{2}}\prod\limits_{j=1}^{2N+1}(\widetilde{\lambda}
- \widetilde{\lambda}_{j})^{\frac{1}{2}}\\& = \mp \zeta^{-N-1}
\prod\limits_{j=1}^{2N+1}(1 -
\widetilde{\lambda}_{j}\zeta)^{\frac{1}{2}} \\& \underset{\zeta
\rightarrow 0}{=} \mp \zeta^{-N-1}\prod\limits_{j=1}^{2N+1} (1 -
\frac{1}{2}\widetilde{\lambda}\zeta
-\frac{1}{8}\widetilde{\lambda}^2 \zeta^2 + O(\zeta^3))\\&
\underset{\zeta \rightarrow 0}{=}\mp\zeta^{-N-1}(1 +
\epsilon_{1}\zeta + \epsilon_{2}\zeta^2 + O(\zeta^3)) \,\,\,\ as
\,\,\ P\rightarrow P_{\infty_{\pm}},
\end{array}
\eqno(4.26)$$ where
$\epsilon_{1}=-\frac{1}{2}\sum\limits_{j=1}^{2N+1}\widetilde{\lambda}_{j},
\,\,\  \epsilon_{2} = \frac{1}{2}\sum\limits_{j < k}
\widetilde{\lambda}_{j}\widetilde{\lambda}_{k} -
\frac{1}{8}(\sum\limits_{j=1}^{2N+1} \widetilde{\lambda}_{j})^2$.
From (3.10) and (2.7), we can derive
$$
\begin{array}{ll}
(\widetilde{\lambda}^{\frac{1}{2}}F)^{-1}&=
\widetilde{\lambda}^{-\frac{1}{2}}\frac{1}{-2u}\prod\limits_{j=1}^N
\frac{1}{\widetilde{\lambda}-\widetilde{u}_{j}} \\&=
-\frac{1}{2u}\zeta^{N + \frac{1}{2}}\prod\limits_{j=1}^N \frac{1}{1
- \widetilde{u}_{j}\zeta}\\& \underset{\zeta \rightarrow
0}{=}-\frac{1}{2u}\zeta^{N+\frac{1}{2}}(1+\sum\limits_{j=1}^N
\widetilde{u}_{j}\zeta + O(\zeta^2)) \,\,\ as \,\ P \rightarrow
P_{\infty_{\pm}}, \end{array} \eqno(4.27)$$
$$
\begin{array}{ll}
V_{12}^{(m)} &= \sum\limits_{j=0}^m
S_{j}^{(2)}\widetilde{\lambda}^{m-j+\frac{1}{2}}\\&
=\zeta^{-\frac{1}{2}}(S_{0}^{(2)}\zeta^{-m} + S_{1}^{(2)} \zeta^{-m
+ 1} + \ldots + S_{m-1}^{(2)}\zeta^{-1} + S_{m}^{(2)})
\end{array} \eqno(4.28)$$
combining (4.4), (3.4), we have:
$$\begin{array}{ll}
& \,\,\,\,\
exp(\int_{t_{m,0}}^{t_{m}}(V_{11}^{(m)}(\widetilde{\lambda}, x_{0},
s) - V_{12}^{(m)}(\widetilde{\lambda}, x_{0}, s)\phi(P, x_{0},
s))ds) \\& = exp(\int_{t_{m,0}}^{t_{m}} (V_{11}^{(m)} -
V_{12}^{(m)}\frac{y +
\widetilde{\lambda}^{\frac{1}{2}}G}{\widetilde{\lambda}^{\frac{1}{2}}F})ds)
\\& = exp(\int_{t_{m,0}}^{t_{m}}(\frac{-y}{\widetilde{\lambda}^{\frac{1}{2}}F}V_{12}^{(m)} +
\frac{1}{2}\frac{F_{t_{m}}}{F})ds) \\&\underset{\zeta \rightarrow
0}{=} exp (\int_{t_{m,0}}^{t_{m}}(\pm\zeta^{-N-1}(1+
O(\zeta))\frac{\zeta^{-\frac{1}{2}}(S_{0}^{(2)}\zeta^{-m} +
S_{1}^{(2)} \zeta^{-m + 1} + \ldots + S_{m-1}^{(2)}\zeta^{-1} +
S_{m}^{(2)})}{\zeta^{-N-\frac{1}{2}}(f_{0}+f_{2}\zeta+ \ldots +
f_{2N}\zeta^N)} + \frac{u_{t_{m}}(x_{0}, s)}{2u(x_{0}, s)} +
O(\zeta))ds)\\& \underset{\zeta \rightarrow 0}{=}
exp(\int_{t_{m,0}}^{t_{m}}(\pm\zeta^{-m-1} + O(\zeta)+
\frac{u_{t_{m}}(x_{0},s)}{2u(x_{0},s)})ds)\\& \underset{\zeta
\rightarrow 0}{=}(\frac{u(x_{0}, t_{m})}{u(x_{0},
t_{m,0})})^{\frac{1}{2}}exp(\pm\zeta^{-m-1}(t_{m}-t_{m,0}) +
O(\zeta)) \,\,\ as P\rightarrow P_{\infty_{\pm}},\\& \underset{\zeta
\rightarrow 0}{=}\left \{
\begin{array}{ll}
exp(\zeta^{-m-1}(t_{m} - t_{m,0}) + O(1)) \,\,\ as \,\,\ P
\rightarrow P_{\infty_{+}},
\\ exp(-\zeta^{-m-1}(t_{m} - t_{m,0}) + O(1)) \,\,\ as \,\,\ P \rightarrow
P_{\infty_{-}},
\end{array}
\right.
\end{array} \eqno(4.29)$$
according to the definition of $\psi_{1}$ in (4.7), (4.25) and
(4.29), we can obtain (4.20). $\Box$

Next, we shall derive the representation of $\phi, \psi_{1},
\psi_{2}, u(x,t_{m}), v(x, t_{m})$ in term of the Riemann theta
function. Let $\omega_{P_{\infty_{+},P_{\infty_{-}}}}^{(3)}$ be the
normalized differential of the third kind holomorphic on
$\mathcal{K}_{N} \backslash \{P_{\infty_{+}}, P_{\infty_{-}}\}$ with
simples at $P_{\infty_{+}}$ and $P_{\infty_{-}}$ and residues $1$
and $-1$ respectively,
$$\omega_{P_{\infty_{+},P_{\infty_{-}}}}^{(3)} =
\frac{1}{y}\prod\limits_{j=1}^N (\widetilde{\lambda} -
\lambda_{j})d\widetilde{\lambda} \underset{\zeta \rightarrow
0}{=}(\pm\frac{1}{2}\zeta^{-1}+O(1))d\zeta \,\,\ as P \rightarrow
P_{\infty_{\pm}}, \eqno(4.30)$$ here the constants
$\{\lambda_{j}|\lambda_{j}\in\mathbb{C}, j=1,\ldots , N\}$ are
uniquely determined by the normalization
$$\int_{a_{j}}\omega_{P_{\infty_{+},P_{\infty_{-}}}}^{(3)}= 0, \,\,\ j=1,\ldots, N, \eqno(4.31)$$
and $\zeta$ in (4.30) denotes the local coordinate $\zeta =
\widetilde{\lambda}^{-1}$ for $P$ near $P_{\infty_{\pm}}$. Moreover,
$$\int_{Q_{0}}^P \omega_{P_{\infty_{+},P_{\infty_{-}}}}^{(3)} \underset{\zeta \rightarrow 0}{=}
\pm(\frac{1}{2}\ln(\zeta)-\ln \omega_{0} + O(\zeta)) \,\,\ as P
\rightarrow P_{\infty_{\pm}}, \eqno(4.32)$$ and $$\int_{Q_{0}}^P
\omega_{P_{\infty_{+},P_{\infty_{-}}}}^{(3)}\underset{\zeta
\rightarrow 0}{=} \ln \omega_{0} + O(\zeta) \,\,\ as P\rightarrow
P_{0}, \eqno(4.33)$$ and $\zeta$ in (4.33) denotes the local
coordinate $\zeta = \widetilde{\lambda}$ for $P$ near $P_{0}$.

Let$\omega_{P_{\infty_{\pm}},r}^{(2)}, \,\ r \ln N_{0}$, be
normalized differentials of the second kind with a unique pole at
$P_{\infty_{\pm}}$, and principal part is $\zeta^{-2-r}d\zeta$ near
$P_{\infty_{\pm}}$, satisfying
$$\int_{a_{j}} \omega_{P_{\infty_{\pm}},r}^{(2)} = 0, j=1,\ldots,N,$$
then we can define $\Omega_{0}^{(2)}$ and $\Omega_{m-1}^{(2)}$ by
$$\Omega_{0}^{(2)} = \omega_{P_{\infty_{+}},0}^{(2)} - \omega_{P_{\infty_{-}},0}^{(2)}, \eqno(4.34)$$
$$\Omega_{m-1}^{(2)} =
\sum\limits_{l=0}^{m-1}\alpha_{m-1-l}(l+1)(\omega_{P_{\infty_{+}},l}^{(2)}-
\omega_{P_{\infty_{-}},l}^{(2)}), \eqno(4.35)$$ where
$\alpha_{m-1-l}, \,\ j=0, \ldots, m-1$ are the integral constants in
(3.8), so we have
$$\int_{a_{j}}\Omega_{0}^{(2)}=0, \int_{a_{j}}\Omega_{m-1}^{(2)}, \,\ j=1, \ldots, N, \eqno(4.36)$$
$$\int_{Q_{0}}^P \Omega_{0}^{(2)} \underset{\zeta \rightarrow 0}{=}
\mp(\frac{1}{2}\zeta^{-1} + e_{0,0} + O(\zeta)) \,\ as P\rightarrow
P_{\infty_{\pm}}, \eqno(4.37)$$
$$\int_{Q_{0}}^P \Omega_{m-1}^{(2)}\underset{\zeta \rightarrow 0}{=}
\mp(\sum\limits_{l=0}^{m-1}\alpha_{m-1-l}\zeta^{-2-l} + e_{m-1,0} +
O(\zeta)) \,\ as P\rightarrow P_{\infty_{\pm}} \eqno(4.38),$$ for
some constants $e_{0,0}, e_{m-1, 0} \in \mathbb{C}$.

If $D_{\underline{\widehat{\widetilde{u}}}(x, t_{m})}$ or
$D_{\underline{\widehat{\widetilde{v}}}(x, t_{m})}$ in (4.5) is
assumed to be nonspecial\cite{Gesztesy-2003}, then according to
Riemann's theorem\cite{Gesztesy-2003,Griffiths-1994}, the definition
and asymptotic properties of the meromorphic function $\phi(P, x,
t_{m})$, $\phi(P, x, t_{m})$ has expressions of the following type:
$$\phi(P, x, t_{m}) = C(x,t_{m})\frac{\theta(P, D_{\underline{\widehat{\widetilde{v}}}(x, t_{m})})}
{\theta(P, D_{\underline{\widehat{\widetilde{u}}}(x,
t_{m})})}exp(\int_{Q_{0}}^P
\omega_{P_{\infty_{+},P_{\infty_{-}}}}^{(3)}), \eqno(4.39)$$ where
$C(x, t_{m})$ is independent of $P \in \mathcal{K}_{N}$.

{\bf Theorem 4.1.} Let $P=(\widetilde{\lambda}, y) \in
\mathcal{K}_{N}\backslash {P_{\infty_{\pm}}, P_{0}}, \,\ (x, t_{m}),
(x_{0}, t_{m,0}) \in \Omega$, where $\Omega \subseteq \mathbb{R}^2$
is open and connected. Suppose $u(\cdot, t_{m}), v(\cdot, t_{m})\in
C^{\infty}(\Omega), u(x, \cdot), v(x, \cdot)\in C^{1}(\Omega), \,\
x\in \mathbb{R}, t_{m} \in \mathbb{R}$, satisfy the equation (1.1),
and assume that $\widetilde{\lambda}_{j}, 1\leq j \leq
2N+2(\widetilde{\lambda}_{2N+2}=0)$ in (3.16) satisfy
$\widetilde{\lambda}_{j} \in \mathbb{C}$ and
$\widetilde{\lambda}_{j} \neq \widetilde{\lambda}_{k}$ for $j \neq
k$. Moreover, suppose that $D_{\underline{\widehat{u}}}$ or
equivalently, $D_{\underline{\widehat{v}}}$, is nonspecial for $(x,
t_{m}) \in \Omega$. Then
$$\phi(P, x, t_{m})=
-\omega_{0}\frac{\theta(P_{\infty_{+}},
D_{\underline{\widehat{\widetilde{u}}}(x, t_{m})})\theta(P,
D_{\underline{\widehat{\widetilde{v}}}(x, t_{m})})}
{\theta(P_{\infty_{+}}, D_{\underline{\widehat{\widetilde{v}}}(x,
t_{m})})\theta(P, D_{\underline{\widehat{\widetilde{u}}}(x,
t_{m})})}exp(\int_{Q_{0}}^P
\omega_{P_{\infty_{+}},P_{\infty_{-}}}^{(3)}), \eqno(4.40)$$
$$
\begin{array}{ll}
&\psi_{1}(P, x, x_{0}, t_{m}, t_{m,0}) =
\frac{\theta(P_{\infty_{+}},
D_{\underline{\widehat{\widetilde{u}}}(x_{0}, t_{m,0})})\theta(P,
D_{\underline{\widehat{\widetilde{u}}}(x,
t_{m})})}{\theta(P_{\infty_{+}},
D_{\underline{\widehat{\widetilde{u}}}(x, t_{m})})\theta(P,
D_{\underline{\widehat{\widetilde{u}}}(x_{0}, t_{m,0})})}
\\& \,\,\,\,\,\,\,\,\,\,\,
\times exp((\int_{Q_{0}}^P \Omega_{0}^{(2)} +
e_{0,0})(x-x_{0})+(\int_{Q_{0}}^P \Omega_{m-1}^{(2)} +
e_{m-1,0})(t_{m}-t_{m,0}))
\end{array}, \eqno(4.41)$$
$$\begin{array}{ll}
&\psi_{2}(P, x, x_{0}, t_{m}, t_{m,0}) = \omega_{0}
\frac{\theta(P_{\infty_{+}},
D_{\underline{\widehat{\widetilde{u}}}(x, t_{m})})\theta(P,
D_{\underline{\widehat{\widetilde{v}}}(x,
t_{m})})\theta(P_{\infty_{+}},
D_{\underline{\widehat{\widetilde{u}}}(x_{0},
t_{m,0})})}{\theta(P_{\infty_{+}},
D_{\underline{\widehat{\widetilde{u}}}(x, t_{m})})\theta(P,
D_{\underline{\widehat{\widetilde{u}}}(x_{0},
t_{m,0})})\theta(P_{\infty_{+}},
D_{\underline{\widehat{\widetilde{u}}}(x, t_{m})})}
\\& \,\,\,\,\,\,\,\,\,\,\,  \times
exp((\int_{Q_{0}}^P \Omega_{0}^{(2)} +
e_{0,0})(x-x_{0})+(\int_{Q_{0}}^P \Omega_{m-1}^{(2)} +
e_{m-1,0})(t_{m}-t_{m,0}))
\\& \,\,\,\,\,\,\,\,\,\,\, \times exp(\int_{Q_{0}}^P \times
\omega_{P_{\infty_{+}},P_{\infty_{-}}}^{(3)})
\end{array}, \eqno(4.42)$$
finally, $u(x, t_{m})$ is of the form
$$u(x, t_{m})= \frac{1}{\omega_{0}^2}\frac{\theta(P_{\infty_{-}},
D_{\underline{\widehat{\widetilde{u}}}(x,
t_{m})})\theta(P_{\infty_{+}},
D_{\underline{\widehat{\widetilde{v}}}(x, t_{m})})}
{\theta(P_{\infty_{-}}, D_{\underline{\widehat{\widetilde{v}}}(x,
t_{m})})\theta(P_{\infty_{+}},
D_{\underline{\widehat{\widetilde{u}}}(x, t_{m})})}, \eqno(4.43)$$
and $v(x, t_{m})$ is determined by
$$u(x, t_{m}) + v(x, t_{m}) + u_{x}(x, t_{m}) - v_{x}(x, t_{m}) =
\frac{\omega_{0}}{2} \partial_{t}\frac{\theta(P_{\infty_{+}},
D_{\underline{\widehat{\widetilde{u}}}(x, t_{m})})\theta(P_{0},
D_{\underline{\widehat{\widetilde{v}}}(x, t_{m})})}
{\theta(P_{\infty_{+}}, D_{\underline{\widehat{\widetilde{v}}}(x,
t_{m})})\theta(P_{0}, D_{\underline{\widehat{\widetilde{u}}}(x,
t_{m})})}. \eqno(4.44)$$

{\bf proof.} We start with the proof of the theta function
representation (4.41).Without loss of generality, it suffices to
treat the special case of (3.8) when $\alpha_{0}=1, \alpha_{k}=0, 1
\leq k \leq N$. First, we assume
$$\widetilde{u}_{j}(x, t_{m}) \neq \widetilde{u}_{k}(x, t_{m}), \,\ for j\neq k,
and \,\ (x, t_{m}) \in \widetilde{\Omega},  \eqno(4.45)$$ for
appropriate $\widetilde{\Omega} \subseteq \Omega$, and define the
right-hand side of (4.41) to be $\widetilde{\psi}_{1}$. In order to
prove $\psi = \widetilde {\psi}$, we investigate the local zeros and
poles of $\psi_{1}$. From (3.10), (3.20), (3.21), (4.4), we have
$$\begin{array}{ll}
\sqrt{\widetilde{\lambda}}u(x^{'}, t_{m})\phi(P, x^{'}, t_{m})
&\underset{P \rightarrow \widehat{\widetilde{u}}_{j}(x^{'},
t_{m})}{=} \sqrt{\widetilde{\lambda}}u(x^{'},
t_{m})\frac{2y(\widehat{\widetilde{u}}_{j}(x^{'},
t_{m}))}{\sqrt{\widetilde{\lambda}}(-2u)\prod\limits_{k=1, k \neq
j}(\widetilde{u}_{j}(x^{'}, t_{m}) - \widetilde{u}_{j}(x^{'},
t_{m}))}\frac{1}{\lambda - \widetilde{u}_{j}(x^{'}, t_{m})}\\&
\underset{P \rightarrow \widehat{\widetilde{u}}_{j}(x^{'},
t_{m})}{=}
\frac{\widetilde{u}_{j,x^{'}}}{\lambda-\widetilde{u}_{j}(x^{'},
t_{m})}\\&\underset{P \rightarrow \widehat{\widetilde{u}}_{j}(x^{'},
t_{m})}{=}-\partial_{x^{'}}\ln(\lambda -\widetilde{u}_{j}(x^{'},
t_{m})) + O(1).
\end{array}
\eqno(4.46)$$ And similarly
$$
V_{12}^{(m)}(\widetilde{\lambda}, x_{0}, s)\phi(P, x_{0}, s)
\underset{P \rightarrow \widehat{\widetilde{u}}_{j}(x_{0}, s)}{=} -
\partial_{s} \ln(\lambda - \widetilde{u}_{j}(x_{0},s)) + O(1), \eqno(4.47)$$
then (4.46) and (4.47) together with (4.7) yields
$$
\psi_{1}(P, x, x_{0}, t_{m}, t_{m,0}) = \left \{
\begin{array}{ll}
(\widetilde{\lambda} - \widetilde{u}_{j}(x,t_{m}))O(1) \,\,\ as
\,\,\ P \rightarrow \widehat{\widetilde{u}}_{j}(x, t_{m}) \neq
\widehat{\widetilde{u}}_{j}(x_{0}, t_{m,0}),
 \\
O(1) \,\,\ as \,\,\,\ P \rightarrow \widehat{\widetilde{u}}_{j}(x,
t_{m}) = \widehat{\widetilde{u}}_{j}(x_{0}, t_{m,0}),
\\ (\widetilde{\lambda} - \widetilde{u}_{j}(x_{0},t_{m,0}))^{-1}O(1) \,\,\ as
\,\,\ P \rightarrow \widehat{\widetilde{u}}_{j}(x_{0}, t_{m,0}) \neq
\widehat{\widetilde{u}}_{j}(x, t_{m}),
\end{array}
\right. \eqno(4.48)$$ where $P=(\widetilde{\lambda}, t_{m}) \in
\mathcal{K}_{N}, (x,t_{m}), (x_{0}, t_{m,0}) \in \widetilde{\Omega}$
and $O(1) \neq 0$. Hence $\psi_{1}$ and $\widetilde{\psi}_{1}$ have
identical zeros and poles on $\mathcal{K}_{N} \backslash
\{P_{\infty_{\pm}}\}$, which are all simple by hypothesis (4.45). It
remains to study the behavior of $\psi_{1}$ and
$\widetilde{\psi}_{1}$ near $P_{\infty_{\pm}}$, by (4.20), (4.37),
(4.38), (4.41), we can easy find that $\psi_{1}$ and
$\widetilde{\psi}_{1}$ share the same singularities and zeros, and
the Riemann-Roch-type uniqueness\cite{Gesztesy-2003} proves that
$\psi_{1} = \widetilde{\psi}_{1}$, hence (4.41) holds subject to
(4.45).

Substituting (4.32), (4.33) into (4.39) and comparing with (4.18)
and (4.19), we obtain
$$u(x, t_{m}) = - \frac{1}{C(x, t_{m})\omega_{0}}
\frac{\theta(P_{\infty_{-}},
D_{\underline{\widehat{\widetilde{u}}}(x,
t_{m})})}{\theta(P_{\infty_{-}},
D_{\underline{\widehat{\widetilde{v}}}(x, t_{m})})}, \,\ C(x, t_{m})
= -\omega_{0} \frac{\theta(P_{\infty_{+}},
D_{\underline{\widehat{\widetilde{u}}}(x,
t_{m})})}{\theta(P_{\infty_{+}},
D_{\underline{\widehat{\widetilde{v}}}(x, t_{m})})}, \eqno(4.49)
$$
according to (4.49), we have (4.43) and (4.40), and $\psi_{2}$ in
(4.42) from $\psi_{2}=-\phi\psi_{1}$. Substituting (4.33) into
(4.39) and comparing with (4.19), together with $C(x, t_{m})$ in
(4.49), we get (4.44).

Hence, we prove this theorem on $\widetilde{\Omega}$. The extension
of all these results from $\widetilde{\Omega}$ to $\Omega$ follows
by continuity of the Abel map and the nonspecial nature of
$D_{\underline{\widehat{\widetilde{u}}}}$ or
$D_{\underline{\widehat{\widetilde{v}}}}$ on $\Omega$. $\Box$

Therefore, the quasi-periodic solution of (1.1) are (4.43) and
(4.44) for $m=1$.

\section{Acknowledgements}
The Project is in part supported by the Natural Science Foundation
of China (Grant No. 11271008), the First-class Discipline of
University in Shanghai and the Shanghai Univ. Leading Academic
Discipline Project (A.13-0101-12-004).


\end{document}